\documentclass[showpacs,amsmath,amssymb,superscriptaddress,aps,a4paper,prl,twocolumn]{revtex4}
\usepackage{graphicx}
\usepackage{dcolumn}
\usepackage{bm}
\usepackage{textcomp}
\usepackage{color}

\begin{document}

\preprint{APS/123-QED}

\title{Coherent detection of electron dephasing}

\author{E.~Strambini}
\email{e.strambini@sns.it}
\affiliation{NEST, Scuola Normale Superiore and CNR-INFM, Piazza dei Cavalieri, 7, I-56126 Pisa, Italy}

\author{L. Chirolli}
\email{luca.chirolli@uni-konstanz.de}
\affiliation{Department of Physics, University of Konstanz, D-78457 Konstanz, Germany}

\author{V. Giovannetti}
\affiliation{NEST, Scuola Normale Superiore and CNR-INFM, Piazza dei Cavalieri, 7, I-56126 Pisa, Italy}

\author{F. Taddei}
\affiliation{NEST, Scuola Normale Superiore and CNR-INFM, Piazza dei Cavalieri, 7, I-56126 Pisa, Italy}

\author{R. Fazio}
\affiliation{NEST, Scuola Normale Superiore and CNR-INFM, Piazza dei Cavalieri, 7, I-56126 Pisa, Italy}

\author{V.~Piazza}
\affiliation{NEST, Scuola Normale Superiore and CNR-INFM, Piazza dei Cavalieri, 7, I-56126 Pisa, Italy}

\author{F. Beltram}
\affiliation{NEST, Scuola Normale Superiore and CNR-INFM, Piazza dei Cavalieri, 7, I-56126 Pisa, Italy}

\date{\today}

\begin{abstract}
We show that an Aharonov-Bohm (AB) ring with asymmetric electron injection can act as a coherent detector of electron dephasing. The presence of a dephasing source in one of the two arms of a moderately-to-highly asymmetric ring changes the response of the system from total reflection to complete transmission while preserving the coherence of the electrons propagating from the ring, even for strong dephasing.
We interpret this phenomenon as an implementation of an interaction-free measurement. 
\end{abstract}


\maketitle

\begin{figure}[t]
\begin{center}
\includegraphics[width=8cm]{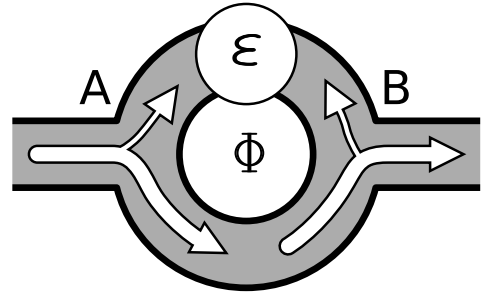}
\caption{Schematic representation of an asymmetric Aharonov-Bohm ring employed to detect the
presence of a dephasing source: in absence of dephasing ($\varepsilon=0$) the 
electron is coherently reflected, while in presence of strong dephasing
($\varepsilon\approx 1$) it is coherently transmitted.\label{Fig1}}
\end{center}
\end{figure}

The observation of quantum coherent phenomena in solid-state systems requires extremely low temperatures and careful design of the experimental setup in order to suppress any source of decoherence.
This represents a major obstacle for the realization of practical coherent electronic devices -- not to mention quantum networks -- as these rely almost completely on the coherent evolution of quantum states, which is in general very effectively destroyed by interactions with the surrounding environment.
In optics, however, it was shown that it is possible to minimize such interactions with a given system, while still being able to gain information about it ~\cite{renninger_158_1960,
elitzur_23_1993, IFM}. These intriguing {\em interaction-free measurement} (IFM) schemes allow, for example, the detection of an absorber while preventing absorption of the probing photons ~\cite{renninger_158_1960, elitzur_23_1993, IFM}.

In this Letter we show that the concept of IFM can be fruitfully extended to electronic quantum-coherent systems.
In this context, differently from the photon case, absorption is not an issue. The target of our IFM scheme is the $coherent$ detection of electron dephasing originating from interactions with the environment and schematized here as external random fluctuating electric fields ~\cite{seelig_64_2001}.
Our IFM-based approach allows one to detect the presence of such noise sources, while preserving electronic-wavefunction coherence.
Specifically, the proposed setup operates as a sort of electronic ``quantum fuse'' that opens or closes a circuit depending on the presence of dephasing noise.
If properly integrated in a multi-lead electronic system (a network), it could in principle be used to steer electron propagation towards regions where dephasing is smaller thereby leading to an increase of the coherence of the electronic flow. 
Decoherence effects play the role of the ``bomb explosion'' discussed in the original Elitzur and Vaidman proposal \cite{elitzur_23_1993} and their impact can be reduced to a negligible level by tuning system parameters, while keeping detection efficiency arbitrarily high. The apparent paradoxical character of this effect arises from a subtle interplay between
destructive interference and state reduction in which the mere presence of the dephasing source in a region of the device prevents the formation of the component of the wave-function that propagates along that path.
This can be achieved by increasing the number of times the probing electrons and the noise source meet, while reducing the probability that at each meeting dephasing occurs. Only a tiny fraction of the electronic wave-function must be diverted to the 
decoherence source: in the asymptotic limit of infinite repetitions of such events the detrimental effect of the interaction (dephasing) is effectively suppressed while it is still possible to reveal the presence of the noise source.
Additionally, as we shall argue in the following, our setup can detect the occurrence of decoherence phenomena with high spatial resolution allowing, for example, to measure very small temperature gradients on the sub-micron scale.

IFMs schemes inspired to the original Mach-Zehnder optical proposal~\cite{elitzur_23_1993}
were discussed for superconducting nano-circuits in Ref.~\cite{PARA} in the context of pulsed-current detection.
The original Mach-Zehnder-based approach can be applied to the electronic case~\cite{chirolli}, however here we choose to refer to a configuration which is more readily suitable for an experimental verification.
We shall investigate a setup based on a device which has been routinely employed for many years in nanoelectronics, namely an Aharonov-Bohm (AB) ring (Fig.~\ref{Fig1}). In our scheme a classical fluctuating electrical field, that acts only on one arm of the ring, randomizes the phase of an electron traveling through it. This field plays the role of the absorber while the loss of coherence mimics the photon absorption of the optical IFM case. 
We assume an asymmetric setup, so that an electron entering the ring from the left (right) lead has a higher
probability of being transmitted through the lower (upper) arm of the ring. As we shall demonstrate in the following, at moderate-to-high asymmetries, an electron injected from the left terminal will preferentially choose one of the two arms (let us suppose for the sake of clarity that the electron chooses the ``lower'' arm, see Fig.~1) and then exit towards the right terminal: the transmission probability of the system is close to one for a broad range of external magnetic fields and gate voltages, except for selected values where the transmission probability shows narrow resonance dips dropping to zero. In these cases the injected electron performs many weak repeated tests of the presence of the dephasing field (supposedly placed in the ``upper'' arm) before reaching the output port: in the absence of the random field the injected electron wavefunction can undergo a coherent evolution that increases the probability of finding it in the ``upper'' arms of the AB ring leading to complete reflection of the electron. On the contrary, when the random field is present, it introduces random phase changes in the tiny portion of electron wavefunction that probes the ``upper'' arm destroying the coherent evolution. The electron remains in the ``lower'' arm, automatically avoiding the path that would lead to dephasing and being almost completely -- and coherently -- transmitted.

We start by introducing the model and analyzing its basic functionality in the idealized zero-temperature case. In the second part of this article we shall discuss finite-temperature effects and determine the threshold below which the effect can be observed.

{\em The model:--}
We consider an asymmetric mesoscopic AB ring, schematized in Fig.~1,
which we characterize in the Landauer-B\"{u}ttiker formulation of quantum transport~\cite{BUTT}. The phase
difference accumulated by the partial wave-functions propagating in the
two arms of the ring can be controlled by means of an external magnetic
field and by gate electrodes, via the magnetic and electric AB effects.
Furthermore, we assume that the asymmetry is such that electron injection into nodes $A$ and $B$
is invariant under a cyclic exchange of the node connectors.
This configuration reproduces, at low magnetic fields, the effects of the
Lorentz force which were studied theoretically~\cite{szafran_72_2005} and realized in the experiments reported in Ref.~\cite{strambini_2009}.
Following Ref.~\cite{strambini_2009}, we parametrize the scattering matrix associated with nodes $A$ and $B$ as follows
\begin{equation}
S_A = \left(\begin{array}{cc}
r_A & \bar{\bf t}_A\\
{\bf t}_A & \bar{\bf r}_A \end{array}\right)
=\left(\begin{array}{ccc}
a & b \cos(\frac{\pi}{2}\gamma) & b \sin(\frac{\pi}{2}\gamma) \\
b \sin(\frac{\pi}{2}\gamma) & a & b \cos(\frac{\pi}{2}\gamma) \\
b \cos(\frac{\pi}{2}\gamma) & b \sin(\frac{\pi}{2}\gamma) & a\\
\end{array}\right) \nonumber
\end{equation}
and 
$S_B=S_A^{\dag}$, with $r_A=a$, ${\bf t}_A$ the $2\times 1$ bottom left block, $\bar{\bf t}_A$ the
$1\times 2$ top right block and $\bar{\bf r}_A$ the remaining $2\times 2$ bottom
right block, with $a=-\sin(\pi\gamma)/(2+\sin(\pi\gamma))$ and $b=\sqrt{1-a^2}$.
The parameter $\gamma$ controls the asymmetry of nodes $A$ and $B$: for $\gamma=0$ or 1 complete asymmetry is achieved, with the electron entering from the left lead being injected totally in the lower or upper arm respectively, whereas for $\gamma=1/2$ injection is symmetric.
Electron propagation in the two arms is described by matrices $S_p(\delta) = e^{i k_F L} \;\mbox{diag}(e^{i\phi/2+i\delta}, e^{-i \phi /2})$, for the transmission from left to right, and $\bar{S}_p(\delta)= e^{i k_F L} \;\mbox{diag}(e^{-i\phi/2+i\delta}, e^{i\phi/2})$, for the transmission from right to left.
Here $\phi$ is the ratio of the magnetic field flux through the ring to the flux quantum, $k_F$ is the Fermi wavenumber, $L$ is the length of the arms and $\delta$ is an additional random phase. In the following we shall set $k_F L=\pi/2$ and anticipate that a different choice does not change qualitatively our findings.

\begin{figure}[t]
\begin{center}
\includegraphics[width=8cm]{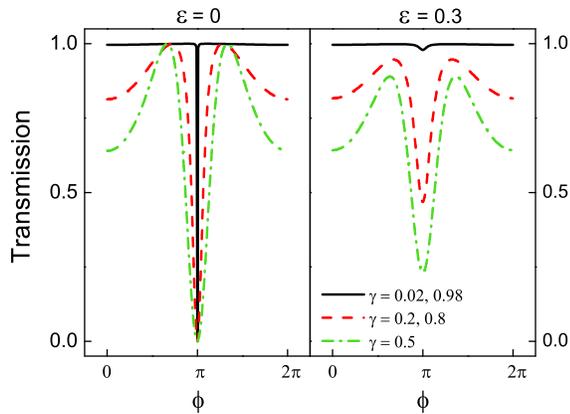}
\caption{Aharonov-Bohm oscillations of the transmission probability $T$ for five
values of the asymmetry $\gamma$, in the case (left) of no
dephasing source and in the case (right) of a dephasing source
with $\varepsilon=0.3$.
The effect of dephasing is strongly enhanced with increasing
asymmetry. \label{Fig2}}
\end{center}
\end{figure}

In the absence of a dephasing source, the transmission amplitude of the ring from the left to the right is given by $t={\bf t}_B(\openone-\Gamma_0)^{-1}S_p(0){\bf t}_A$, where $\Gamma_0=S_p(0)\bar{\bf r}_A\bar{S}_p(0){\bf r}_B$ (the bar indicates right-to-left processes).
As shown in the left panel of Fig.~\ref{Fig2}, at zero temperature the system shows characteristic Aharonov-Bohm
oscillations of the transmission probability $T=|t|^2$ with a well-defined zero-valued minimum at the working point $\phi = \pi$.
Such minimum becomes narrower as the asymmetry is increased, i.e. when $\gamma$ approaches 0 or 1.
In this case, when $\phi\neq\pi$, injected electrons are preferentially transmitted through the lower ($\gamma$ close to 0) or upper ($\gamma$ close to 1) arm so that $T=1$ and negligible interference takes place.
When $\phi=\pi$, however, the marked destructive interference survives despite the very small probability for an electron to choose the upper ($\gamma$ close to 0) or lower ($\gamma$ close to 1) arms, giving rise to the narrow dip in the transmission.
This situation resembles the one realized in multi-round concatenated interferometers where optical IFM is observed~\cite{IFM}. There, the asymmetry is introduced by choosing interferometer beam-splitters with a reflection (or transmission) probability close to 1.

Let us now assumethat a fluctuating external field (dephasing source) is placed in the upper arm of the ring.
To account for it we define the partial transmission amplitude of order $N$ as
$t_N={\bf t}_B\sum_{n=0}^N\prod_{j=0}^n\Gamma_{(n-j)}S^{(0)}_p{\bf t}_A$,
where $\Gamma_{(j)}\equiv\Gamma(\delta_j,\delta_j') =
S_p(\delta_j)\bar{\bf r}_A\bar{S}_p(\delta_j'){\bf r}_B$ depends
on two random phases $\delta_j$ and $\delta'_j$, and $S_p^{(0)}\equiv S_p(\delta_0)$.
We then choose the random phases from a distribution
$g_{\varepsilon}(\delta)$ of zero mean and width $2\pi\varepsilon$ and compute the averaged
partial transmission probability as $\langle t_N^* t_N \rangle_{\delta}$,
where $\langle\ldots\rangle_{\delta}=
\int d\boldsymbol{\delta}g_{\varepsilon}(\boldsymbol{\delta})\ldots$, and
$g_{\varepsilon}(\boldsymbol{\delta})=g_{\varepsilon}(\delta_0)\ldots g_{\varepsilon}(\delta_{2N})$~\cite{not}.
It can be shown that the following recursive relation holds:
$\langle t_N^*t_N\rangle_{\delta}=\langle t_{N-1}^*t_{N-1}
\rangle_{\delta}+\Xi_N$.
By iterating the procedure, the averaged transmission probability $\langle T\rangle_{\delta} = \lim_{N\rightarrow\infty} \langle t_N^* t_N\rangle_{\delta}$
can be written as
$\langle T\rangle_{\delta}=\sum_{N=0}^{\infty}\Xi_N$.
In order to compute this limit we introduce the Pauli matrix vector $\boldsymbol{\sigma}=
(\sigma_0,\sigma_1,\sigma_2,\sigma_3)^T$, with $\sigma_0=\openone$,
and define the following decoherence matrix
\begin{equation}\label{Eq:DecohMap}
{\cal Q}_{ij}=\frac{1}{2}\int d\boldsymbol{\delta}~
g_{\varepsilon}(\boldsymbol{\delta}){\rm Tr}\left[
\Gamma^{\dag}(\boldsymbol{\delta})\sigma_i\Gamma(\boldsymbol{\delta})
\sigma_j\right],
\end{equation}
which allows us to
perform the average over the random phase as a
matrix product.
Similarly we define $\Gamma_{\rm av}=\int d\boldsymbol{\delta}~g_{\varepsilon}(\boldsymbol{\delta})
\Gamma(\boldsymbol{\delta})$,
and the decoherence map ${\cal P}$ with entries

\begin{equation}
{\cal P}_{ij}=\frac{1}{2}\int d\delta g_{\varepsilon}(\delta){\rm Tr}\left[
S_p^{\dag}(\delta)\sigma_i S_p(\delta)\sigma_j\right],
\end{equation}
that describes the average over the random phase in $S_p^{(0)}$. 
$\Xi_N$ can now be concisely written as:
\begin{equation}
\Xi_N={\bf t}_A^{\dag}\left({\bf p}_B\cdot {\cal Q}^N
+\sum_{k=1}^N{\bf p}_k\cdot{\cal Q}^{N-k}\right)
\cdot{\cal P}\cdot\boldsymbol{\sigma}~
{\bf t}_A,
\end{equation}
with the vector $({\bf p}_k)_i=\frac{1}{2}\left[{\rm Tr}({\bf t}^{\dag}_B
{\bf t}_B\Gamma_{\rm av}^k\sigma_i)+c.c\right]$.
By writing ${\bf p}_k={\rm Re}[\lambda_1^k\Lambda_1+\lambda_2^k\Lambda_2]
\cdot{\bf p}_B$, with $\lambda_i$ the eigenvalues of $\Gamma_{\rm av}$,
$U$ the matrix of the eigenvectors of $\Gamma_{\rm av}$, and
$(\Lambda_i)_{jk}=(U\sigma_j\sigma_kU^{-1})_{ii}$,
that satisfy $(\Lambda_1+\Lambda_2)/2=\openone$, we can perform
the sum on $N$ and obtain:
\begin{eqnarray}\label{Eq:Dephased-T}
\langle T\rangle_{\delta}&=&{\bf t}^{\dag}_A~{\bf p}_B\cdot
({\cal T}-\openone)\cdot(\openone-{\cal Q})^{-1}
\cdot{\cal P}\cdot\boldsymbol{\sigma}~{\bf t}_A,
\end{eqnarray}
with ${\cal T}$ being a 4x4 matrix defined by ${\cal T}=\sum_{i=1,2}{\rm Re}[\frac{1}{1-\lambda_i}\Lambda_i^T].$

\begin{figure}[t!]
\begin{center}
\includegraphics[width=8cm]{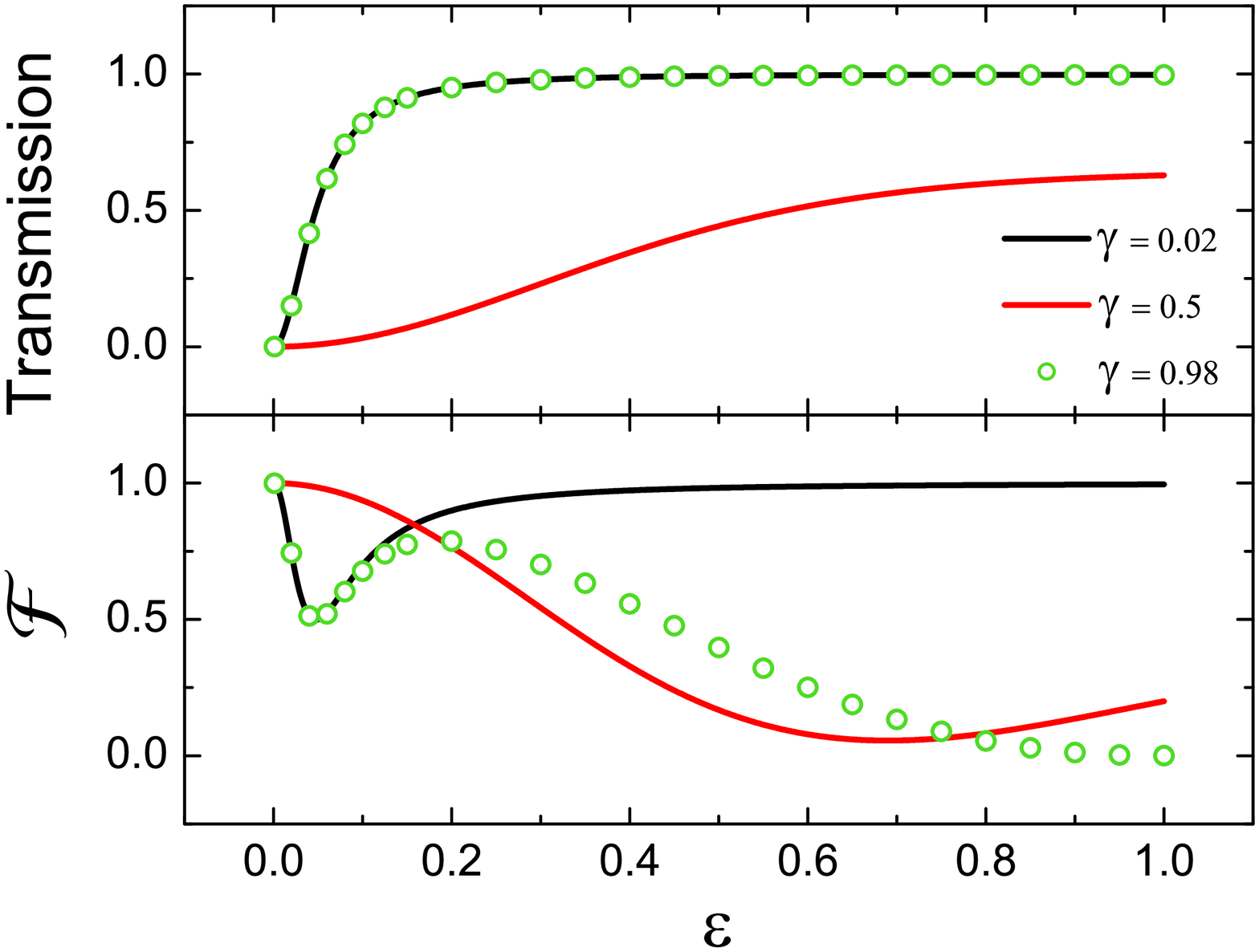}
\caption{Effect of the dephasing on the transmission probability (up) and {\em coherence function} (down) as a function of $\varepsilon$ at resonance ($\phi$ = $\pi$) for three values of the asymmetry parameter $\gamma$.
\label{Fig3}}
\end{center}
\end{figure}

The effect of a dephasing source, placed in the upper arm of the ring, on the transmission probability $T$ is presented in the right panel of Fig.~\ref{Fig2} for $\varepsilon=0.3$.
As expected, a reduction of the visibility of the oscillations is found which is more pronounced for the narrow dips.
This can be viewed as a ``which-path detection'' \cite{PATH}, whereby by means of the dephasing process and the consequent suppression of the destructive interference occurring at $\phi=\pi$ the ``environment'' acquires the information that the electron propagated along the upper arm.
In particular, for large asymmetries, corresponding to $\gamma$ values close to 0 or 1 (the cases $\gamma=0.02$ and $\gamma=0.98$ are shown in the figure,) the presence of the dephasing source leads to an almost complete suppression of the narrow transmission dip. 
This effect is further highlighted by the upper panel of Fig.~\ref{Fig3} which shows the transmission probability at the working point $\phi=\pi$ as a function of $\varepsilon$ for different $\gamma$ values. $T$ is found to increase when $\varepsilon$ increases, reaching almost total transmission for large asymmetries.
It should be noted that the transmission probability as a function of $\varepsilon$ is invariant under the transformation $\gamma \rightarrow 1 - \gamma$, that corresponds to invert the asymmetry of the ring, or, equivalently, to move the position of the noise source from one arm to the other. This symmetry implies that while measuring $T$ provides a faithful estimation of the noise intensity parameter $\varepsilon$, it does not allow one to determine on which arm the decoherence source is acting.


The fact that the presence (absence) of dephasing in the upper arm is signaled by the nearly complete reflection (transmission) of the injected electrons does not constitute, as such, an IFM.
On the contrary, IFM of the dephasing source requires that the ``outgoing signal'' is not degraded: in the present case, electrons should preserve their phase coherence once transmitted or reflected by the ring. 
We shall demonstrate that in our device this is indeed occurring by evaluating the overall phase coherence by means of the following {\em coherence function}
${\cal F}=|\langle t\rangle_{\delta}|^2+|\langle r\rangle_{\delta}|^2$, where
$\langle t\rangle_{\delta}$ ($\langle r\rangle_{\delta}$) is the averaged transmission (reflection) amplitude.
${\cal F}$ takes values between 0 (complete loss of coherence) and 1 (coherence fully preserved,
since in this case $|\langle t\rangle_{\delta}|^2 = T$ and $|\langle r\rangle_{\delta}|^2= R = 1 -T$).
The two quantities $\langle t\rangle_{\delta}$ and $\langle r\rangle_{\delta}$ measure the coherence of the transmitted and reflected electrons, respectively, since they are proportional to the interference terms of such electrons with a reference, coherent, signal.
An estimate of ${\cal F}$ can be achieved in practice by inserting the ring in one arm of a larger loop, and measuring the visibility of the oscillations as a function of an additional phase shift occurring in the other arm of the loop (further details will be published elsewhere~\cite{chirolli}).

In the lower panel of Fig.~\ref{Fig3}, ${\cal F}$ is plotted as a function of the width $\varepsilon$ of the distribution of random phases, for three different values of $\gamma$.
Let us first consider the case $\gamma=0.02$. For very small values of $\varepsilon$, ${\cal F}$ decreases from unity as a result of the degradation of coherence. 
For large values of $\varepsilon$, the dephasing of the tiny portion of the wavefunction probing the upper arm prevents the occurrence of destructive interference and allows full, {\em coherent}, transmission through the lower arm yielding ${\cal F}\simeq 1$.
This can be understood as due to the quantum Zeno effect so that, for large $\varepsilon$, the electrons are ``repeatedly measured'' in the upper arm by the ``environment'' \cite{sudarshan}. For $\gamma$ close to zero, the outcome of this measurement will be negative with a very high probability (i.e. the electron is found in the lower arm) preserving coherence.
An interplay between these two effects occurs for intermediate values of $\varepsilon$ giving rise to a minimum in ${\cal F}$.
For $\gamma$ close to 1 ($\gamma=0.98$ in the figure) electrons are mostly injected in the upper arm of the ring. For small values of $\varepsilon$ the situation is analogous to the case $\gamma=0.02$, the behavior of ${\cal F}$ being actually the same, the role of the noise source being the partial suppression of the destructive interference and consequent degradation of the coherence (the position of the noise source is virtually unimportant).
Large values of $\varepsilon$ yields a drop of ${\cal F}$ to zero, since the complete suppression of the destructive interference is accompanied by a likewise complete loss of coherence due to the fact that most of the electrons visit the noise source.
For a symmetric ring ($\gamma=0.5$), as expected, ${\cal F}$ decreases smoothly up to $\varepsilon\simeq 0.7$ where it takes a small but finite value before inverting its trend. 
At large asymmetries, a measurement of ${\cal F}$ would allow, not only to detect the presence of the noise source, but also to determine on which arm it is acting, something which can be of paramount importance for the accurate settings of thermoelectric measurements at the nanoscale.

{\em Finite temperature effects:--} In order to assess the effect of a finite temperature we study the extent of degradation
of the AB oscillations induced by thermal phase averaging.
In this case the differential conductance is defined as
$G=\frac{e^2}{h}\int dE\left(-\partial f/\partial E\right)|t(E)|^2$ and since
we assume that matrices $S_A$ and $S_B$ negligibly depend on energy close to the Fermi level $E_F$, the energy dependence in the transmission amplitude stems only from the
wave number $k(E)=\frac{1}{\hbar}\sqrt{2m^*(E+E_F)}$, with $m^*$ the effective
mass of the electron. For a ring with arms of equal length $L$, in the absence of dephasing, the energy dependence of the transmission probability can be explicitly written: 
$|t(E)|^2=|{\bf t}_B(\openone-e^{i\pi E/(2E_F)}\Gamma_0)^{-1}S_p{\bf t}_A|^2$.
Operatively, we can set a threshold temperature $T_t\sim 10^{-4}~E_F/k_B$ which corresponds to a
reduction of the dip smaller than 1\%.
The typical value $E_F\approx 5~{\rm meV}$ gives $T_t\approx 10~{\rm mK}$, a temperature within experimental reach.
We note that a small temperature difference between the arms of the ring could be detected as a IFM as well.

In conclusion we demonstrated that IFM can be implemented for electrons in the solid state
where a dephasing source plays the role of the absorber of the optical counterpart.
IFM for electrons has different properties from its optical analogue. In particular not only can detection of a dephasing source in one of the arms of the ring be achieved without degrading the outgoing electrons, but also loss of coherence or its preservation allow the determination of the position of the dephasing source with high spatial resolution. 
We have focused on a very simple system, namely an asymmetric ring, but the physics remains unchanged for different implementations.
The present proposal provides a test of non trivial quantum mechanical effects at the mesoscopic level and may find 
useful applications in quantum information, e.g.  allowing for fault-tolerant electronic circuitry.

The Authors thank G.~Burkard for discussions and comments.
This work was supported by the Italian Ministry of University and Research under the FIRB
IDEAS project ESQUI.
L.~C.~acknowledges funding from the DFG within SPP 1285 ``Spintronics''
and from the Swiss SNF via grant n0. PP02-106310. V.~P.~acknowledges CNR-INFM for funding through the SEED
Program.


\begin{thebibliography} {99}
\bibitem{not} From now on we shall assume a uniform distribution centered around $\delta=0$.
\end{thebibliography}
\end{document}